\documentclass[prl,10pt,twocolumn,groupedaddress,floatfix,showpacs]{revtex4-1}

\usepackage[utf8]{inputenc}  
\usepackage[T1]{fontenc}     
\usepackage[british]{babel}  
\usepackage[sc,osf]{mathpazo}
\usepackage[scaled=0.86]{berasans}  
\usepackage[colorlinks=true, allcolors=blue, urlcolor=blue]{hyperref}  
\usepackage{graphicx} 
\usepackage[babel]{microtype}  
\usepackage{amsmath,amssymb,amsthm,bm,amsfonts,mathrsfs,bbm} 

\usepackage{xspace}  
\usepackage{pgfplots}
\usepackage{xcolor,colortbl}
\usepackage{array}
\usepackage{bigstrut}

\newcommand{\N}{\mathbb{N}}

\newcommand{\Q}{\mathbb{Q}}
\newcommand{\E}{\mathbb{E}}

\newcommand{\tr}{\text{tr}}

\newcommand{\be}{\begin{equation}}
\newcommand{\ee}{\end{equation}}
\newcommand{\bea}{\begin{eqnarray}}
\newcommand{\eea}{\end{eqnarray}}
\newcommand{\bes}{\begin{equation*}}
\newcommand{\ees}{\end{equation*}}
\newcommand{\beas}{\begin{eqnarray*}}
	\newcommand{\eeas}{\end{eqnarray*}}

\def\E{\mathrm{E}}
\def\N{\mathcal{N}}

\def\tr{\mathrm{tr}}

\def\L{\mathcal{L}}
\def\Q{\mathcal{Q}}
\def\NL{\mathcal{NS}}

\newtheorem*{thm*}{Theorem}

\newtheorem*{lem*}{Lemma}

\newtheorem*{lipschitzLem*}{Lemma \ref{lipschitz}}
\newtheorem*{lipschitzCubeLem*}{Lemma \ref{lipschitzCube}}
\newtheorem*{pgmNearlyOptimalThm*}{Theorem \ref{pgmNearlyOptimal}}

\begin{document}

\title{ Non-Local Network Coding in Interference Channels }


\author{Jiyoung Yun$^{1}$}
\email{jiyoungyun@kaist.ac.kr}

\author{Ashutosh Rai$^{1,2}$}
\email{ashutosh.rai@savba.sk}

\author{Joonwoo Bae$^{1}$}
\email{joonwoo.bae@kaist.ac.kr}

\affiliation{$^{1}$School of Electrical Engineering, Korea Advanced Institute of Science and Technology (KAIST), 291 Daehak-ro, Yuseong-gu, Daejeon 34141, Republic of Korea,}

\affiliation{$^{2}$Institute of Physics, Slovak Academy of Sciences, 845 11 Bratislava, Slovakia.}


\begin{abstract}

In a network, a channel introduces correlations to the parties that aim to establish a communication protocol. We present a framework of nonlocal network coding by exploiting a Bell scenario and show the
usefulness of nonlocal and quantum resources in network coding. Two-sender and two-receiver interference channels are considered, for which network coding is characterized by two-input and
four-outcome Bell scenarios. It is shown that nonsignaling correlations lead to strictly higher channel capacities than quantum correlations in general. This also holds true for quantum and local correlations: network coding with quantum resources shows a strictly higher channel capacity than local ones. It turns out, however, that more nonlocality does not necessarily imply a higher channel capacity. The framework can be generally applied to network communication protocols. 
\end{abstract}


\maketitle






A network generates correlations. The parties that aim to establish a communication protocol via a network have to deal with interventions due to the correlations. {\it Network coding} presents a framework to devise codewords for reliable communication in a network via cooperation of the parties \cite{Gamal_and_Kim}. Messages are chosen by senders, mapped to codewords by network coding, and then transmitted to receivers through a network channel. 

A multipartite Bell scenario presents a natural network framework that maps input bits to outputs bits. Inputs are chosen by the parties randomly and independently and a Bell scenario may generate correlations among outcomes, classified into local and non-local ones. The parties with shared randomness only are compatible with local correlations but non-local ones. 



It turns out that the non-local correlation characterized by the Popescu-Rohrlich (PR) box \cite{PRbox} is useful for enhancing a channel capacity in a network channel such as an interference channel \cite{QueckShor2005}. The result has been extended to a multiple access channel by generalizing to non-local games \cite{smith}. Entanglement has been asserted as a useful resource for the network coding. Along the line, it is shown that computing channel capacities is a difficult problem, e.g., NP-Hard \cite{smith}. Moreover, a general formulation of how to exploit non-local correlations in network coding is lacking. Consequently, little is known about network coding and the usefulness of non-classical correlations in network communications, apart from particular cases of maximally entangled states or maximally non-local probabilities \cite{QueckShor2005}. It is also worth mentioning that nonclassical correlations, such as entanglement, the non-locality, steering, etc., are generally inequivalent resources with each other \cite{r0, r1,r2,r3}.




In this work, we establish a Bell scenario as a framework of {\it non-local network coding} that applies non-signaling correlations beyond shared randomness to the preparation of codewords for reliable network communication. We in particular consider two-input and two-output interference channels, to show that almost all non-local (quantum) correlations are more useful for higher channel capacities than quantum (local) correlations. It is also shown that more non-local correlations do not necessarily imply to a higher channel capacity, i.e., the non-locality is not a general resource that enhances a network protocol. The framework is also useful  to construct non-local polytopes containing the set of quantum correlations. Our results can be generally applied to other network channels when non-local resources are available in network coding. 

\begin{figure}[t]
	\begin{center}
		\includegraphics[angle=0, width=0.48\textwidth]{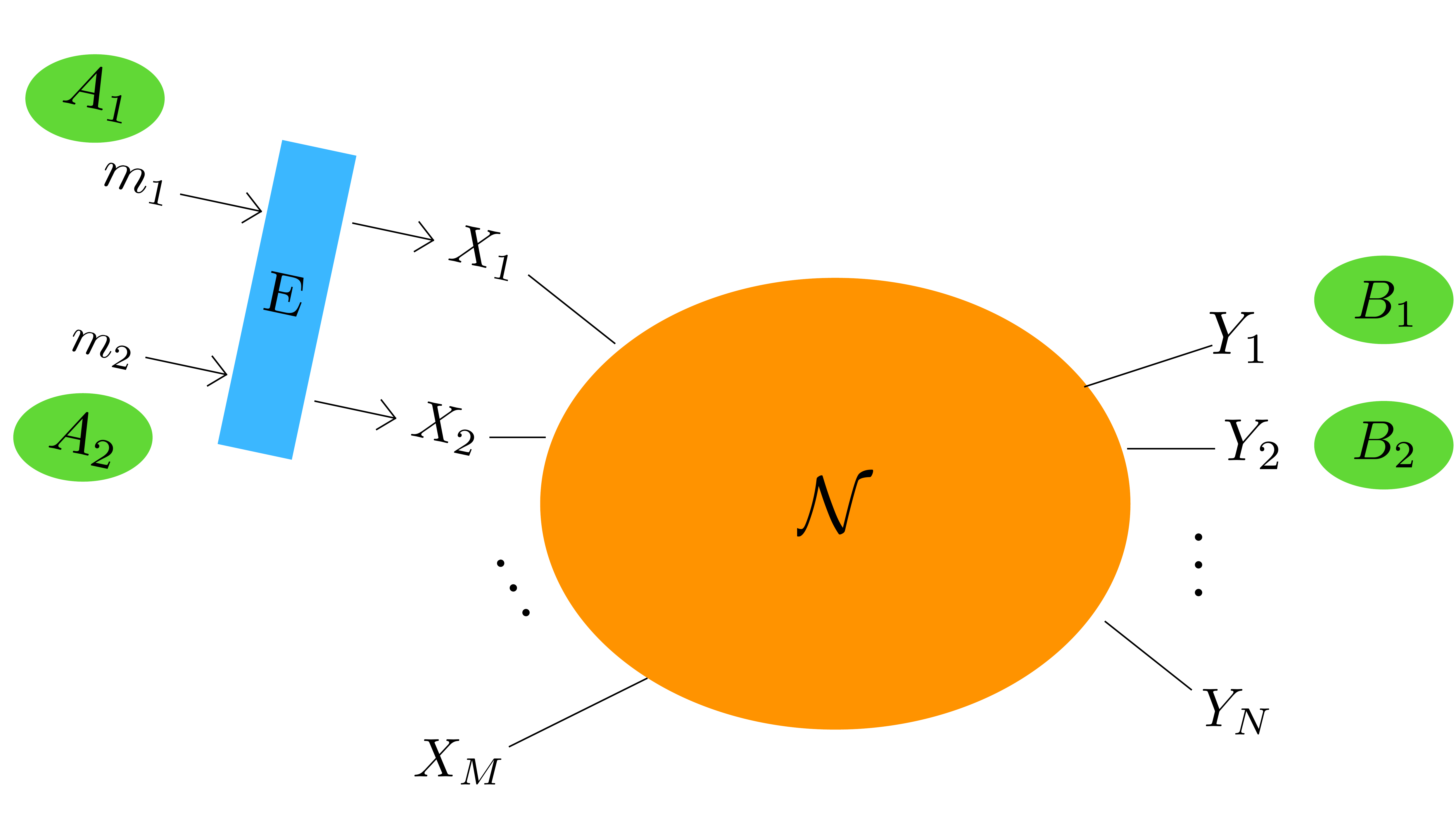}
		\caption{ In a network of $m+n$ parties, two senders $A_1$ and $A_2$ choose two inputs $m_1$ and $m_2$ as the messages to deliver to parties $B_1$ and $B_2$. The senders exploit a Bell scenario for network coding. For a channel considered throughout in Eq. (\ref{channel}), network coding corresponds to a CGLMP scenario, see the main text.} \label{fig1}
	\end{center}	
\end{figure}

Let us begin with a network channel of many inputs ($m$) and many outputs ($n$), denoted by, see also Fig. \ref{fig1}, 
\bea 
\N : (X_1,X_2 \cdots, X_m) \rightarrow ( Y_1, Y_2,\cdots, Y_n). \nonumber
\eea
A channel is characterized by its conditional probability $P_{\N }(Y_1  Y_2 \cdots  Y_n | X_1  X_2 \cdots  X_m)$. For instance, a point-to-point channel $ X_i \rightarrow Y_j $ can be found by a marginal probability $P_{\N }(Y_j   | X_i  )$. We are interested in a two-input and two-output interference channel throughout, which is characterized by a joint probability $P_{\N }(Y_1 Y_2|X_1 X_2  )$.

The interference channels of our interest here are those satisfying $I(X_1;Y_2) \neq 0$  or $I(X_2;Y_1)\neq 0$ for some input distribution $P(X_1X_2)$, where $I$ denotes the mutual information. These channels are known as incompatible or non-separate interference channels, which in fact show the effects of interference. It has been shown that an interference channel $\N$ is incompatible if its conditional probability $P_{\N}(Y_1 Y_2\vert X_1 X_2)$ satisfies the following ~\cite{Sato},
\bea
&& P_{\N}(Y_1\vert X_1 X_2)\neq P_{\N}(Y_1\vert X_1 X_{2}^{'}) ~\mathrm{ for}~ X_2 \neq X_{2}^{'} \nonumber\\
&\mathrm{ or}~~&P_{\N}(Y_1 \vert X_1 X_2)\neq P_{\N}(Y_1\vert X_{1}^{'} X_2) ~\mathrm{ for}~ X_1\neq X_{1}^{'}. \nonumber
\eea
This shows that the no-signaling conditions are not fulfilled in an incompatible network channel. Correlations of any kind can be generated between the parties communicating via the interference channel.


Network coding in a two-input and two-output interference channel works as follows. Let $A_1$ and $A_2$ denote random variables of two senders and $B_1$ and $B_2$ of two receivers. Network coding is implemented by a mapping $\E : (A_1,A_2) \rightarrow (X_1,X_2)$, where we have $X_1,X_2 = \{0,1 \}^d$ in general. Let $m_1 \in A_1$ and $m_2 \in A_2$ denote message bits chosen by the senders, respectively. The map can be characterized by its joint probability $P_{\E} (X_1 X_2 | A_1 A_2)$. Together with an encoding scheme $\E$ to an interference channel $\N$, the transmission from two senders $(A_1, A_2)$ to two outputs $(Y_1,Y_2)$ is characterized by a joint probability in the following,
\bea
&&P_{\N\circ\E} (Y_1 Y_2\vert A_1 A_2) =\nonumber \\
&& \sum_{ X_1 X_2 }  P_{\N} ( Y_1 Y_2 \vert X_1 X_2) 
 P_{\E} ( X_1  X_2  \vert A_1 A_2).  \label{eq:condp}
\eea
With a decoding scheme $(Y_1, Y_2)\rightarrow (B_1,B_2)$, the goal is to find the sum capacity defined as follows, 
\bea
&& C^{(R)} (\N) =\max_{\E \in R} ~C_{\E }^{(R)} (\N),~ \label{eq:enc}\\
&& \mathrm{where} ~C_{\mathrm{E}}^{(R)} (\N) = I_{\E} (A_1 :  B_1) + I_{\E}(A_2 : B_2). \label{eq:sum} 
\eea
The maximization in Eq. (\ref{eq:enc}) runs over encoding schemes $\E$ with an available resource $R$. The sum rate $C_{\mathrm{E}}^{(R)} (\N)$ has been defined in Eq. (\ref{eq:sum}), where the mutual information with an encoding scheme $\E$ is denoted by $I_{\E}$. Note that a decoding does not increase the mutual information since it would correspond to a mapping between sets of alphabets of an equal size. W.l.o.g., it suffices to optimize an encoding $\E$ to find the sum capacity. 

One of the important properties of interference channels is that for any encoding or decoding schemes, the sum rate depends only on the marginal distributions $P_{\N}(Y_1\vert X_1 X_2)$ and $P_{\N}(Y_2\vert X_1 X_2)$. This can define equivalence classes of interference channels in terms of marginal distributions: namely, channels having the same marginal distributions are equivalent.

We in particular consider a class of two-sender and two-receiver interference channels, where $X_i = (x_{i1}, x_{i2})$ for $i=1,2$ and $x_{i1},x_{i2}, Y_1, Y_2 \in \{ 0,1\}$, see Fig. \ref{fig1}. The channels are characterized by joint probabilities with two parameters as follows, 
\begin{eqnarray}
\begin{matrix} 
P_{\N} (Y_{1}=x_{11},Y_{2}=x_{21})&=&p \\ 
P_{\N} (Y_{1}\neq x_{11},~ Y_{2} \neq x_{21})&=&1-p
\end{matrix} \nonumber 
\bigg\} &~&\mathrm{if}~ x_{12}\oplus x_{22}=x_{11}x_{21},\\[2pt]
\begin{matrix} 
P_{\N}(Y_{1}=x_{11},Y_{2}=x_{21})&=&q \\ 
P_{\N}(Y_{1}\neq x_{11}, Y_{2}\neq x_{21})&=&1-q
\end{matrix} 
\bigg\}  &~&\mathrm{otherwise}. \label{channel}
\end{eqnarray}  
where $p,q\in[0,1]$. These are obtained by generalizing the interference channel introduced in Ref. \cite{QueckShor2005}, which can be found as the case that $p=1$ and $q=0$. Note also that the channels with $p=q\in[0,1]$ are excluded since they are not incompatible.

For the channels in Eq. (\ref{channel}), network coding works as a mapping from two inputs to four outcomes, 
\bea
\E:   (m_1,m_2)  \mapsto ((x_{11},x_{12}),(x_{21},x_{22}) )  \label{eq:netco}
\eea
where two inputs $ m_1, m_2 \in \{0,1 \}$ are messages chosen by space-like separated and non-communicating parties $A_1$ and $A_2$ respectively. The coding scheme is equivalent to a Bell scenario~\cite{bellnonlocality} of two space-like separated parties, i.e., two senders, who choose from a set of two measurements where each measurement has four possible outcomes. This can be referred to as a Collins-Gisin-Linden-Massar-Popescu (CGLMP) scenario with four outcomes \cite{CGLMP}. The CGLMP scenario with $2$ inputs $x,y\in\{0,1\}$ and $4$ outcomes $a,b\in\{0,1,2,3\}$ corresponds to network coding $\E$ in Eq. (\ref{eq:netco}) by relating the outcomes as follows, $\{0\rightarrow (0,0),~1\rightarrow (0,1),~2\rightarrow (1,0),~3\rightarrow (1,1)\}.$

As network coding in Eq. (\ref{eq:netco}) is identical to a CGLMP scenario, encoding schemes $\E$ can be therefore classified into local, quantum, and non-local network coding, according to correlations $P_{\E} (X_1 X_2 | A_1 A_2)$ in Eq. (\ref{eq:condp}). That is, resources available in network coding are classified accordingly. First, {\it local network coding} is referred when local correlations are exploited in the encoding: the codewords are prepared by shared randomness only
\begin{equation}
P_{\E \in \mathcal{L}} (X_1X_2\vert A_1 A_2 )=\sum_{ i,j=1}^{16} p_{ij } D_i(X_1\vert A_1) D_j(X_2\vert A_2), \label{local}
\end{equation}
where $D_{i}$ and $D_{j}$ for $i,j\in \{1,2,...,16\}$ are all possible deterministic functions $A_i\rightarrow X_i$ for $i=1,2$ and $p_{ij}$ denotes the shared randomness between two senders $A_1$ and $A_2$. Next, {\it quantum network coding} can be characterized by conditional probabilities compatible with quantum theory as follows, 
\begin{equation}
P_{\E \in \mathcal{Q}} (X_1X_2\vert A_1 A_2 )= \tr (\rho_{A_1A_2}~\Pi^{A_1}_{X_1}\otimes\Pi^{A_2}_{X_2}) \label{quantum}
\end{equation}
for some quantum state $\rho_{A_1A_2}$ and measurements $\Pi^{A_1}_{X_1}$ and $\Pi^{A_2}_{X_2}$. This means two parties prepare codewords by sharing quantum states and measurements on them. Then, \emph{non-local network coding} is referred when two senders have access to all non-signaling probabilities in the encoding,
\bea
&& P_{\E \in \mathcal{NS}} (X_1X_2\vert A_1 A_2 )~ \mathrm{such ~that~} \label{eq:nonlocal} \\
&& P_{\E \in \mathcal{NS}} (X_1\vert A_1A_2) =  P_{\E \in \mathcal{NS}}  (X_1\vert A_1 A_{2}^{'}) ~\mathrm{for}~A_{2} \neq A_{2}^{'}\nonumber \\
&& P_{\E \in \mathcal{NS}} (X_2\vert A_1A_2) = P_{\E \in \mathcal{NS}}  (X_2\vert A_{1}^{'} A_2) ~\mathrm{for}~A_{1} \neq A_{1}^{'}. \nonumber 
\eea
The sets of local ($\L$), quantum ($\Q$), and non-signaling ($\NL$) probabilities are convex and strict hierarchical $\mathcal{L}\subsetneq \mathcal{Q}\subsetneq \mathcal{NS}$. That is, non-signaling (quantum) correlations contain quantum (local) ones. 


In fact, the aforementioned strict hierarchy structures the sum rate in Eq. (\ref{eq:sum}). To see this, we recall a useful lemma in information theory: the mutual information $I (X;Y)$ is a convex function of conditional probabilities $p(y|x)$ for a fixed distribution $p(x)$ \cite{Cover_and_Thomas}. By fixing $P(A_1 A_2) =1/4$ for $A_1, A_2 \in \{0,1 \}$, we have that both $I_{\E} (Y_1 |A_1) $ and $I_{\E} (Y_2 |A_2) $ in the sum rate in Eq.~(\ref{eq:sum}) from a joint probability $P_{\N\circ\E} (Y_1 Y_2  A_1 A_2)$ are convex functions of conditional probabilities  $P_{\E} ( X_1   X_2  \vert A_1  A_2)$ in Eq.~(\ref{eq:condp}). For two encoding schemes $\E_1$ and $\E_2$, their convex mixtures $\E=\lambda \E_1 + (1-\lambda) \E_2$ can be characterized by conditional probabilities as follows, $P_{\E} ( X_1   X_2  \vert A_1 A_2) = \lambda P_{\E_1} ( X_1   X_2  \vert A_1  A_2) + (1-\lambda )P_{\E_2} ( X_1   X_2  \vert A_1  A_2)$. Due to the convexity of the mutual information {\it w.r.t.} conditional probabilities, it follows that 
\bea
C_{\E}^{(R)} (\N) & \leq & \lambda C_{\E_1}^{(R)} (\N) + (1-\lambda) C_{\E_2}^{(R)} (\N)  \label{eq:convex} 
\eea
 This in fact leads to a simplification in the computation of the sum capacity: consequently, it suffices to consider all extreme encodings in the optimization in Eq. (\ref{eq:sum}). \\

	{\bf Lemma}. The sum rate $C_{\E}^{(R)} (\N)$ in Eq. (\ref{eq:sum}) is a convex function of encoding schemes $\E$. The sum capacity corresponds to a maximal sum rate over extreme elements in the set of encoding schemes. \\

In what follows, we show that the strict hierarchy for channel capacities according to the resources in network coding, namely 
\bea
C^{(\NL)} (\N) > C^{(\Q)}(\N)  >C^{(\L)}(\N), \label{eq:result}
\eea
for channels in Eq. (\ref{channel}) for almost all $p,q\in[0,1]$ . From Lemma, the task is to find extreme encoding schemes in the set of local, quantum, and non-signaling correlations. To this end, we are going to exploit the CGLMP scenario. 

 	The convex geometry of the sets $\mathcal{L},~\mathcal{Q},~\mbox{and}~\mathcal{NS}$ has been analyzed for the CGLMP scenario \cite{Barrett2005}. The local polytope is identified by the convex hull of all local deterministic points, which are also finite. It suffices to explore the finite vertices to compute the sum capacity. The non-signaling polytope is given by the convex hull of all local and nonlocal points. Note that the nonlocal properties of the vertices are invariant under \emph{local reversible relabeling} of inputs and outputs of individual parties in a Bell scenario \cite{Barrett2005,Jones_Masanes}, see also Appendix. The set of vertices up to local reversible relabelings gives the full characterization of the non-signaling polytope. 
	
All local verticies are characterized by reversible local relabelings of a representative local deterministic vertex in the following,
\bea
P(a,b|x,y)=
\begin{cases}
	1 & \text{if }~a=0~ \mbox{and}~b=0 \\[2pt]
	0 & \text{otherwise }.\label{localvertex}
\end{cases}
\eea
All non-local vertices are obtained by reversible local relabelings of a set of three representative non-local vertices for $k =2,3,4$, respectively, as follows, 
\bea
P(a,b|x,y)=
\begin{cases}
	\frac{1}{k} & \text{if }~ (b-a)~ \text{mod}~ k = xy ~\\ 
	&\text{for}~ a,b\in \{0,\cdots, k-1\},\\[2pt]
	0 & \text{otherwise }. \label{nonlocalvertex}
\end{cases}
\eea
Let $V_j$ denote a set of vertices generated by local reversible relabelings of the polytopes having maximal probabilities $j \in\{1,\frac{1}{2},\frac{1}{3},\frac{1}{4}\}$ in Eqs. (\ref{localvertex}) and (\ref{nonlocalvertex}), respectively. The vertices are exploited to find an optimal coding for the channel capacity, see Eqs. (\ref{eq:enc}) and (\ref{eq:netco}).

The sum capacity with local network coding $C^{(\L)} (\N)$ can be obtained by exploring all local deterministic vertices characterized by $V_1$ in Eq. (\ref{localvertex}). Applying all local reversible relabelings, there are $256$ local deterministic vertices. The details are shown in Appendix. The sum capacity is found as follows,
\begin{eqnarray}
&C^{(\L)} (\N) &=\max~\{ 1- h(p), 1-h(q), f_L (p,q), f_L (q,p)\}  \nonumber \\
\mathrm{with}&f_L(p,q)&=2~h (\frac{2+p-q}{4} ) -h ( \frac{p+q}{2} )-h(p),~~~~~~ \label{eq:lcap}
\end{eqnarray}
where $h$ denotes the binary entropy. 

The sum capacity with non-local network coding $C^{(\NL)} (\N)$ can be obtained by applying all reversible local relabelings to the three sets of the non-local verticies in Eq.~(\ref{nonlocalvertex}). The non-signaling polytope $V_{\mathcal{NS}}=V_1 \cup V_{\frac{1}{2}}\cup V_{\frac{1}{3}}\cup V_{\frac{1}{4}}$ consists of $204160$ vertices. The sum capacity is obtained as follows, 
\begin{eqnarray}
&  C^{(\NL)} (\N) &=\max~\{ 2( 1- h(p)), 2(1-h(q)) \}. \label{nscap}
\end{eqnarray}
From the sum capacities in Eqs. (\ref{eq:lcap}) and (\ref{nscap}), it holds that $C^{(\NL)} (\N) > C^{(\L)} (\N) $ for all $p,q\in[0,1]$. For instance, for $(p,q) =(1,0)$ we have $C^{(\L)} (\N)=1$ and $C^{(\NL)} (\N) =2$. Note that the maximal capacity $2$ is obtained if and only if vertices of $V_{\frac{1}{2}}$ are applied in network coding in Eq. (\ref{eq:nonlocal}). 

For the computation of the sum capacity with quantum resources, it is essential to have the characterization of the quantum set $\mathcal{Q}$. This is, however, a hard problem classified as NP-Hard \cite{NPA1,NPA2}. Then, our strategy here is to construct two convex polytopes $\Q_{LB}$ and $\Q_{UB}$ such that $\Q_{LB} \subset \Q \subset \Q_{UB}$. The goal is to find bounds for the quantum sum capacity, i.e., $C^{(\Q_{LB})} \leq C^{( \Q) } \leq C^{( \Q_{UB} )}$. Note also that the sets $\Q_{LB}$ and $\Q_{UB}$ are constructed to have finite number of vertices to make the computation feasible.

\begin{figure}[t]
	\begin{center}
		\includegraphics[angle=0, width=0.48\textwidth]{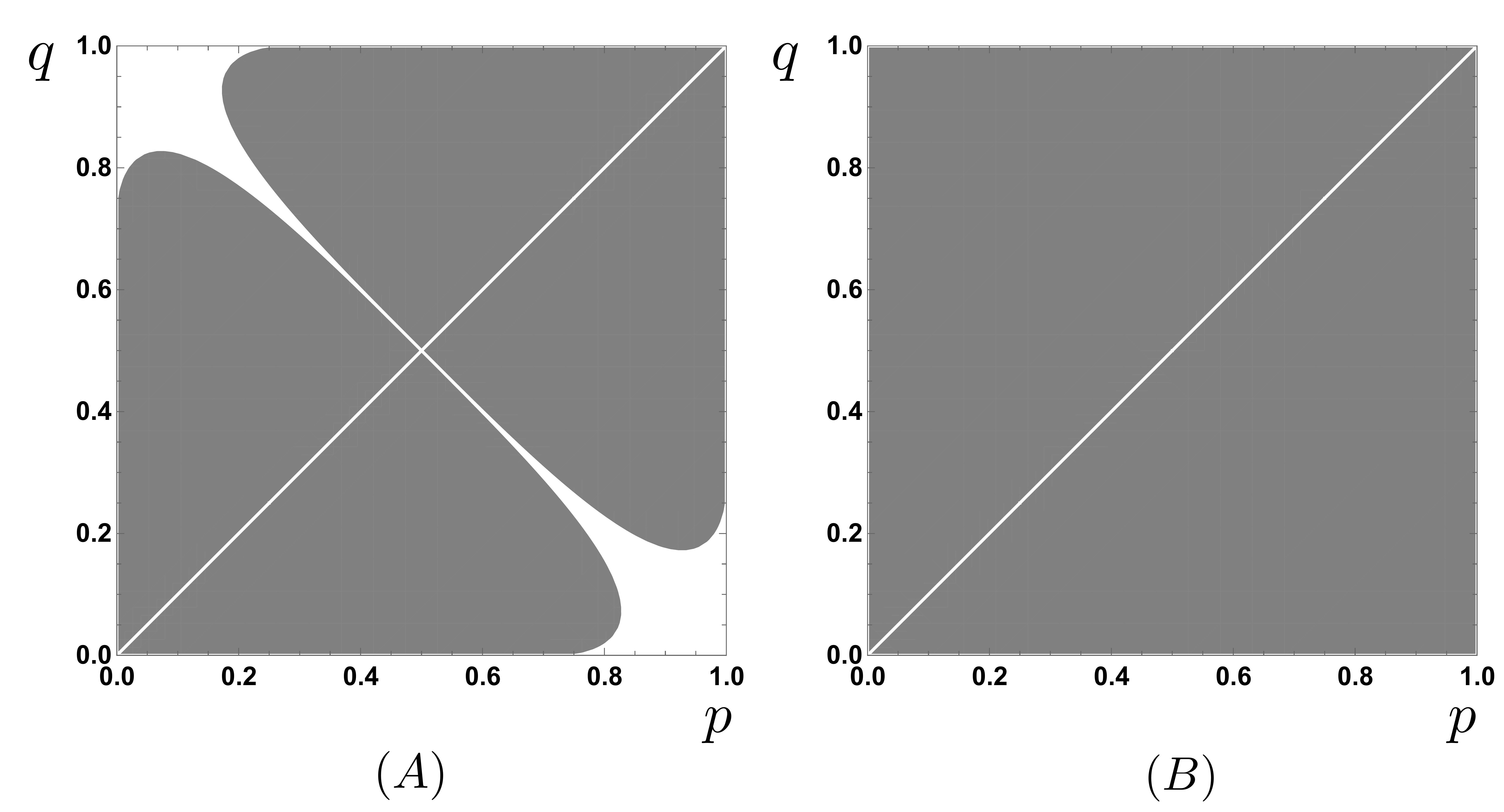}
		\caption{ In the gray area, it holds that (A) $C^{(\Q)} (\N) \geq C^{(\Q_{LB})} (\N) > C^{(\L)} (\N)$ from Eqs. (\ref{eq:lcap}) and (\ref{eq:qclb}), and (B) $C^{(\Q)} (\N) \leq C^{(\Q_{UB})} (\N) < C^{(\NL)} (\N)$ from Eqs. (\ref{nscap}) and (\ref{eq:qub}). For almost all $p,q \in[0,1]$, non-local network coding is more useful than quantum network coding, which is more useful than local network coding.} \label{almost}
	\end{center}	
\end{figure}

For lower bounds, we make use of a Bell scenario as follows. The maximally entangled state $\vert \Psi \rangle_{A_1A_2}= \frac{1}{\sqrt{2}}(\vert 01\rangle-\vert 10\rangle)$ is shared by two senders. For $i=1,2$, let $m_i$ denote a message bit that a party $A_i$ wants to send to $B_i$. The party $A_1$ applies a measurement $\sigma_z$ for $m_1=0$ and $\sigma_x$ for  $m_1=1$. The other $A_2$ performs a measurement $-\frac{1}{\sqrt{2}}(\sigma_x+\sigma_z)$ for $m_2=0$ and $\frac{1}{\sqrt{2}}(\sigma_x-\sigma_z)$ for $m_2=1$. Let $a_1,a_2\in\{0,1 \}$ denote the outcomes respectively. By maximizing the sum rate over encodings given by all deterministic functions, i.e., $E: (m_1,m_2) \mapsto (DF_{1}(m_1,a_1),~DF_{2}(m_2,a_2))$ for deterministic functions 
$ DF_i~:\{00,01,10,11\} \mapsto \{00,01,10,11\}$ for $i=1,2$, we find that

\begin{eqnarray}
	&& C^{ (\Q_{LB} )} (\N)=\max~\{ 1-h(p), 1-h(q), f_{LB}(p,q) , f_{LB}(q,p) \},\nonumber \\
	&&\mathrm{where}~ f_{LB}(p,q)=2-2 h ( \frac{2+\sqrt{2}}{4}p+\frac{2-\sqrt{2}}{4}q ).\label{eq:qclb} 
\end{eqnarray}
Since the Bell experiment is a quantum strategy, a lower bound to the quantum capacity is shown.

To compute an upper bound to the quantum capacity, we construct a polytope that contains the quantum set. While the tight characterization when the number of vertices is fixed is highly non-trivial, our strategy is to exclude those vertices giving the maximal capacity $C^{(R\in \NL)}=2$ in the non-signaling polytope $V_{\NL}$, where the maximal capacity is from those vertices equivalent to $V_{\frac{1}{2}}$ in Eq. (\ref{nonlocalvertex}). The polytope constructed by remaining vertices consists of the quantum set. By exploring the vertices of the polytope, we have an upper bound,
\begin{eqnarray}
	&& C^{ (\Q_{UB})} (\N)=\max~\{ g  (p,q), g (q,p), f_{UB}(p,q) , f_{UB}(q,p) \},\nonumber \\
        &&\mathrm{where} ~g (p,q) =1+h ( \frac{2+p}{5} )-\frac{1}{2} h (\frac{1+3p}{5} )-\frac{3}{2} h(p),\nonumber \\ 
	&& \mathrm{ and}~  f_{UB} (p,q) =2 h ( \frac{5+p-q}{10} ) - h ( \frac{4p+q}{5} ) - h( p ). ~~~~\label{eq:qub}
\end{eqnarray}
With the upper and lower bounds to the quantum capacity, one can find that the strict hierarchy in Eq. (\ref{eq:result}) holds true for almost all $p,q\in[0,1]$, see Fig. \ref{almost}. Thus, it is shown that non-local (quantum) correlations are more useful than quantum (local) ones in network coding. 

We now investigate the relation of the non-locality and the channel capacity, and show that the more non-locality does not necessarily imply a higher channel capacity. As a measure for the non-locality, we consider the variational distance from the local polytope $\mathcal{L}$. The measure has been devised from the perspective of resource theories of the non-locality \cite{brito}. For the considered scenario, the measure is proportional to the violation of the CGLMP inequality \cite{CGLMP}. 

A simplified form of a two-input and four-outcome CGLMP inequality has been shown, that for all local probabilities $P_{\E\in\L} (ab|xy)$,
\begin{eqnarray}
&& B_4 (P_{\E\in\L} (a b |x y)) := ~P(a\leq b\vert 00)+P(a\geq b\vert 01)+ \nonumber \\
&&P(a\geq b\vert 10)+P(a<b\vert 11)-3 ~\leq~ 0 \label{cglmp}
\end{eqnarray}
for $  x,y\in \{0,1\}$ and $a,b\in \{0,1,2,3\}$ \cite{cglmp2, cglmp3}. The measure can be written as, $D_{NL}[P(ab|xy)] = \max \{0,B_4 (P(ab | xy)) \}$. All different forms of CGLMP inequalities can be derived from Eq. (\ref{cglmp}) by reversible local relabelings. For instance, one can consider that local inputs and outputs remain the same for $x,y=0$ whereas for $x,y=1$, local outputs $a$ and $b$ are relabeled to $(a+2)~\mbox{mod}~4$ and $b$ to $(b+2)~\mbox{mod}~4$. An equivalent CGLMP inequality can be obtained as follows,
\bea
&& \widetilde{B}_4  (P_{\E\in\L} (ab|xy)) :=P(a\leq b\vert 00)+P(a\geq b\oplus 2 \vert 01)+ \nonumber \\
&& P(a\oplus 2 \geq b\vert 10)+P(a\oplus 2<b\oplus 2\vert 11)-3 ~\leq~ 0, \label{cglmpnew}
\eea
where $r\oplus s\equiv(r+s)~\mbox{mod}~4$.

\begin{table}[]
\begin{tabular}{c||cccc }
 & $v_{\frac{1}{2} }$  & $v_{\frac{1}{3} }$ &  $v_{\frac{1}{4} }$ &  \\
 \hline\hline
$ $ &   &   &   &  \\
$\widetilde{B}_4 (v)$ ~~ & ~~1/2~~ & ~ ~2/3~~ & ~~ 3/4~~ &  \\
\hline
$ $ &   &   &   &  \\
$C_{\E=v }^{(\NL)} (\N)$ ~~& ~2~ & ~1.4570~ & ~1.4322 ~&  
\end{tabular}
\caption{Three non-local vertices $v_{\frac{1}{2}}, v_{\frac{1}{3}}, v_{\frac{1}{4}} $ are compared. The non-locality is measured in the first row, and the sum rate in the second. The vertex $v_{\frac{1}{4}}$ is the most non-local but the least useful for network coding, whereas the least non-local one $v_{\frac{1}{2}}$ is the most useful for network coding. }\label{table}
\end{table}

To compare the non-locality and the channel capacity, let us consider three vertices denoted by $v_{\frac{1}{2}}$, $v_{\frac{1}{3}}$, and $v_{\frac{1}{4}}$ in the non-signaling polytope, see Appendix for details. In Table \ref{table}, the non-locality measured by $\widetilde{B}_4 (v)$ and the sum rate $C_{\E=v}^{(\NL)} (\N) $ are compared for each vertex $v\in\{ v_{\frac{1}{2}},  v_{\frac{1}{3}} , v_{\frac{1}{4}}\}$. It is shown that the most non-local correlation is the least useful for network coding, and also that the least non-local one is the most useful for a network communication. This shows that non-local correlations are not a general resource that enhances a network protocol.

In conclusion, we have established a framework of network coding with a Bell scenario. Local, quantum, and non-local network coding schemes are characterized accordingly. We have constructed equivalent Bell inequalities by exploiting the technique of local reversible relabelings to solve the optimization in network coding. On the technical side, our method can be used to construct non-local polytopes that contain the quantum set.

It is shown that non-local resources are strictly more useful than quantum resources, which are strictly more useful than local resources. We have also shown that the non-locality is not a general resource for enhancing network communications. More non-locality does not necessarily leads to a higher rate in network communication. 


Our results shed a new light to understand network information theory. The results find that non-local correlations are generally useful in network communication. The framework we presented here with a Bell scenario can be applied to other network channels in general. To compute a channel capacity over the quantum set, it is asked to develop an efficient method of constructing non-local polytopes for optimizations in network coding. We leave it an open question to seek theoretical tools for the purpose. In general, network coding can be found by characterizing constrained non-local polytopes in the probability space. Finally, our work paves a way to develop multipartite Bell scenarios for network communication. In future investigations, it would be interesting to apply multipartite Bell scenarios to multiple-input multiple-output (MIMO) network channels.

\begin{acknowledgments} 
This work is supported by the National Research
Foundation of Korea (Grant No. 2019M3E4A1080001),
an Institute of Information and Communications Tech-
nology Promotion (IITP) grant funded by the Korean
Government (MSIP) (Grant No. 2019-0-00831, EQGIS)
and the Information Technology Research Center Program
(IITP-2020-2018-0-01402). AR also acknowledges funding and support from VEGA Project No. 2/0136/19
and the Slovak Academy of Sciences.
\end{acknowledgments}

\appendix

\section*{APPENDICES}

\section{I.~~Algorithm for implementing all local reversible relabellings in a four outcome CGLMP scenario \label{apdxI}}

We consider a Bell scenario of two parties, say Alice and Bob. Let $x\in \mathcal{X}$ and $a\in \mathcal{A}$ denote an input and an output of Alice, respectively, and $y \in \mathcal{Y}$ and $b\in \mathcal{B}$ an input and an output of Bob, respectively. A probability space of vectors as follows
\bea 
v = (P(0,0 | 0,0), P(0,1 | 0,0),\cdots ) \nonumber
\eea
can be defined with probabilities \{~$P(a,b\vert x,y): x\in \mathcal{X}, ~a\in \mathcal{A}, ~y\in \mathcal{Y},~ b\in \mathcal{B}$~\}. In a probability space, a Bell inequality corresponds to a hyperplane that distinguishes a non-local one from local ones, where a Bell inequality is satisfied by all local probabilities but some non-local ones. 

A local reversible relabeling (LRR) is a method of relabeling inputs and outputs locally such that probabilities can be transformed with each other. Note that an LRR does not generate a non-local correlation and is also invertible. An LRR defines an equivalence class in a probability space, i.e., two probability vectors are equivalent if there is an LRR that transforms one to the other and vice versa. 

Consequently, a set of non-local probabilities can be transformed to an equivalent set under an LRR, according to which the corresponding Bell inequality can be generated. A number of Bell inequalities generated by LRRs are equivalent in the sense that those non-local probabilities detected by the Bell inequalities are equivalent under LRRs. I.e., non-local properties are invariant under an LRR. In what follows, we show how to perform an LRR on probabilities in practice to find equivalent non-local probabilities and to generate  equivalent Bell inequalities.

\subsection{(a).~~Local Reversible Relabellings  \label{apdxIa}}

A {\it Local Reversible Relabeling} is a reversible mapping of input and output (given the input) which can be implemented locally by space-like separated parties. An LRR can be characterized by a transformation $\{P(a,b\vert x,y)\}\rightarrowtail \{P(\tilde{a},\tilde{b}\vert \tilde{x},\tilde{y})\}$ according to some functions $\tilde{x}=f^{A}_{i}(x)$, ~$\tilde{a}=f^{A}_{o}(x,a)$, ~$\tilde{y}=f^{B}_{i}(y)$, ~$\tilde{b}=f^{B}_{o}(y,b)$ such that all the functions $f^{A}_{i},~f^{A}_{o},~f^{B}_{i},~f^{B}_{o}$ are bijective, i.e., invertible. It is clear that the functions are local transformations in the sense that the constraints of a Bell experiment are fulfilled. 

In the CGLMP scenario with $4$ outcomes for each of the two parties, LRRs can be implemented by all permutations of the input sets $\mathcal{X}=\{0,1\}~\mbox{and}~\mathcal{Y}=\{0,1\}$, and all permutations of the ouputs $\mathcal{A}=\{0,1,2,3\}~\mbox{and}~\mathcal{B}=\{0,1,2,3\}$ for a given local input. Thus, the total number of LRRs is given by  $(2!\times 4!\times 4!)\times (2!\times 4!\times 4!)$.

\subsection{(b).~~Algorithm  \label{apdxIb}}

Let us say $\vec{P}_{in}=\{P_{in}(a,b\vert x,y)\}$, where $x,y \in \{0,1\}$ and $a,b\in \{0,1,2,3\}$, is an initial probability vector. Then, on applying an LRR to  $\vec{P}_{in}$, we obtain another probability vector  $\vec{P}_{out}=\{P_{out}(a,b\vert x,y)\}$. Here, in what follows, for the four outcome CGLMP scenario, we are considering all possible LRRs. We note that, depending on the initial probability point $\vec{P}_{in}$, many LRRs may lead to the same final probability point  $\vec{P}_{out}$. First we implement all the LRRs on a given initial probability vector to generate a multiset consisting of $N$ final probability vectors. Then we delete all duplicates in the  multiset and obtain a set of all distinct probability vectors resulting from applying all possible LRRs to the initial probability vector $\vec{P}_{in}$.

(i) \emph{Define:} Permutation functions on set of inputs $I=\{0,1\}$: ~$PER^{I}[r](\cdot)$ for $r\in\{1,2\}$.

(ii) \emph{Define:} Permutation functions on set of outputs $O=\{0,1,2,3\}$: ~$PER^{O}[s](\cdot)$ for $s\in\{1,2,...,24\}$.

(iii) \emph{Define:} LRR functions $F[i,j,k_0,l_0,k_1,l_1](\cdot)$, for $i,j\in\{1,2\}$ and $k_0,l_0,k_1,l_1\in\{1,2,...,24\}$, as follows 
\begin{itemize}
	\item \emph{Alice}: 
	\begin{itemize}
		\item applies $PER^{I}[i](\cdot)$ to input $x$,
		\item when input $x=0$, applies $PER^{O}[k_0](\cdot)$ to output $a$,
		\item when input $x=1$, applies $PER^{O}[k_1](\cdot)$ to output $a$.
	\end{itemize}
	
	\item \emph{Bob}: 
	\begin{itemize}
		\item applies $PER^{I}[j](\cdot)$ to input $y$,
		\item when input $y=0$, applies $PER^{O}[l_0](\cdot)$ to output $b$,
		\item when input $y=1$, applies $PER^{O}[l_1](\cdot)$ to output $b$.
	\end{itemize}
	\item Total number of LRR functions $F[i,j,k_0,l_0,k_1,l_1](\cdot) $ are 
	$$N=(2!\times 4!\times 4!)\times (2!\times 4!\times 4!)$$
\end{itemize}

\begin{enumerate}
	\item Input: probability vector $\vec{P}_{in}=\{\{P_{in}(a,b\vert x,y)\}: x,y \in \{0,1\} ~\mbox{and}~a,b \in \{0,1,2,3\}\}.$
	
	\item Compute: MultiSet of output probability vectors, by applying $N$ LRR functions:
	\begin{eqnarray}
	M_S&=&\{\vec{P}_{out}=F[i,j,k_0,l_0,k_1,l_1](\vec{P_{in}}): i,j\in\{1,2\},\nonumber \\ 
	&~& \mbox{and}~k_0,l_0,k_1,l_1\in\{1,2,...,24\}\}.\nonumber
	\end{eqnarray}
	\item Output: ~$S= \mbox{DeleteDuplicates}[M_S].$
	
\end{enumerate}

\section{II.~~All vertices in the four outcome CGLMP scenario  \label{apdxII}}

The algorithm in the above for applying LRRs is implemented in MATHEMATICA by considering as input(s) the four representative vertices in a two-input and four-outcome CGLMP scenario. 

\begin{itemize}
	\item Local representative vertex 
	\bea
	P(a,b|x,y)=
	\begin{cases}
		1 & \text{if }~a=0~ \mbox{and}~b=0 \\[2pt]
		0 & \text{otherwise },\label{A1}
	\end{cases}
	\eea
	gives the set of all local deterministic vertices $V_1$ consisting of $256$ elements.
	
	\item Nonlocal representative vertex with probabilities either $\frac{1}{2}$ or $0$
	\bea
	P(a,b|x,y)=
	\begin{cases}
		\frac{1}{2} & \text{if }~ (b-a)~ \text{mod}~ 2 = xy ~\\ 
		&\text{for}~ a,b\in \{0,1\},\\[2pt]
		0 & \text{otherwise }, \label{A2}
	\end{cases}
	\eea
	gives the set all nonlocal vertices $V_{\frac{1}{2}}$ of its type, and it consists $10368$ elements.
	\item Nonlocal representative vertex with probabilities either $\frac{1}{3}$ or $0$
	\bea
	P(a,b|x,y)=
	\begin{cases}
		\frac{1}{3} & \text{if }~ (b-a)~ \text{mod}~ 3 = xy ~\\ 
		&\text{for}~ a,b\in \{0,1,2\},\\[2pt]
		0 & \text{otherwise }, \label{A3}
	\end{cases}
	\eea
	gives the set all nonlocal vertices $V_{\frac{1}{3}}$ of its type, and it consists $110592$ elements.
	
	\item Nonlocal representative vertex with probabilities either $\frac{1}{4}$ or $0$
	\bea
	P(a,b|x,y)=
	\begin{cases}
		\frac{1}{4} & \text{if }~ (b-a)~ \text{mod}~ 4 = xy ~\\ 
		&\text{for}~ a,b\in \{0,1,2,3\},\\[2pt]
		0 & \text{otherwise }, \label{A4}
	\end{cases}
	\eea
	gives the set all nonlocal vertices $V_{\frac{1}{4}}$ of its type, and it consists $82944$ elements.
	
\end{itemize}

\section{III.~~Classical capacity  \label{apdxIII}}

The channel capacity with local network coding is obtained by generating the set all local deterministic vertices $V_1$. This is done by applying all LRRs to the local deterministic vertex given by Eq.~(\ref{A1}); the number of local deterministic vertices are $256$. Next, with the help of MATHEMATICA, we computed the sum-rate $C_{\E}^{(\L)}$ on all vertices in $V_1$. These are basically different functions of $p$ and $q$. Then by noticing that many of the vertices give same sum rate functions, we delete all the duplicates to get only $9$ different functions. Since for finding an expression for capacity we need to take maximum over all these functions, we filter and eliminate all the functions which are always less then or equal to a smaller subset of the nine functions.  

We find that the formula for classical capacity can be expressed as follows,
\begin{eqnarray}
&C^{(\L)} (\N) &=\max~\{ 1- h(p), 1-h(q), f_L (p,q), f_L (q,p)\}  \nonumber \\
\mathrm{with}&f_L(p,q)&=2~h (\frac{2+p-q}{4} ) -h ( \frac{p+q}{2} )-h(p),~~~~~~ \label{eq:lcap}
\end{eqnarray}
where, here and henceforth, $h$ denotes the binary entropy.

\section{IV.~~No-signaling capacity  \label{apdxIV}}

For deriving the no-signaling capacity, we first generate the sets of nonlocal vertices $V_{\frac{1}{2}}$, $V_{\frac{1}{3}}$, and $V_{\frac{1}{4}}$ by applying all LRRs to the three respective nonlocal vertex given by Eqs.~(\ref{A2}), (\ref{A3}), (\ref{A4}). The total number of vertices in the three sets $V_{\frac{1}{2}}$, $V_{\frac{1}{3}}$, and $V_{\frac{1}{4}}$ are respectively $10368$, $110592$, and $82944$ . Thus we could get the set of all no-signaling vertices $V_{\mathcal{NS}}=V_1 \cup V_{\frac{1}{2}}\cup V_{\frac{1}{3}}\cup V_{\frac{1}{4}}$ consisting of $204160$ total number of vertices. Then, with the help of MATHEMATICA, we compute  the total sum rate functions on all these vertices and by observing that many of the vertices give same sum-rate functions we delete all the duplicates. Finally, since we are interested in maximum over all these functions,  we filtered out most of these function to obtain that the formula for no-signaling capacity can be expressed as,
\begin{eqnarray}
&  C^{(\NL)} (\N) &=\max~\{ 2( 1- h(p)), 2(1-h(q)) \}. \label{nscap}
\end{eqnarray}

\section{ V.~~Bounds to Quantum capacity  \label{apdxV}}

The quantum set of correlations $\mathcal{Q}$ is a convex set but not a polytope which makes computation of quantum capacity a hard problem. In fact, it is hard to characterize the exact boundary of the set $\mathcal{Q}$: this has been classified as an NP-Hard problem. Therefore, one can derive some lower and upper bounds to the quantum capacity by approximating the quantum set from inside and outside with the help of numerical tools and well-defined geometries of local, quantum, and non-local correlations. In particular, we exploit two polytopes, one within the quantum set for a lower bound, and the other containing the quantum set for the upper bound. In what follows, we adopt this approach for deriving suitable lower and upper bounds on quantum capacities.

\subsection{(a).~~Lower bound  \label{apdxVa}}

The computation of a lower bound (Eq. (15) in the main text) in the main text is obtained from all possible encodings $\mathrm{E}(X_1,X_2\vert m_1,m_2)$, that correspond to local deterministic functions. The probability distribution that gives the maximum Bell-CHSH violation (i.e., Tsirelson's point for the CHSH scenario) is obtained. Then, all deterministic mappings are applied to input and output bits of the Tsirelson correlation in order to prepare inputs to the channel, which corresponds to an encoding. The deterministic mappings are computed via MATHEMATICA , and the sum-rate is computed for all these encodings. Then, a minimal set from these functions is selected such that the best possible lower bound can be achieved for the considered set of quantum protocols. The expression for a lower bound on quantum capacities resulting from this protocol is as follows,	
\begin{eqnarray}
&& C^{ (\Q_{LB} )} (\N)=\max~\{ 1-h(p), 1-h(q), f_{LB}(p,q) , f_{LB}(q,p) \},\nonumber \\
&&\mathrm{where}~ f_{LB}(p,q)=2-2 h ( \frac{2+\sqrt{2}}{4}p+\frac{2-\sqrt{2}}{4}q ).\label{eq:qclb} 
\end{eqnarray}

It is also worth to mention other approaches we have made to compute a lower bound. The initial distributions in the following are considered: (i) quantum boundary points giving maximal violation of tilted-CHSH inequalities \cite{r1}, and (ii) points on the  Tirelson-Landau-Masanes (TLM) boundary of the quantum set in CHSH scenario \cite{a1,a2,a3}. Although the attempts with (i) and (ii) have found some improvements in the lower bound for the channel parameters $p$ and $q$, it turns out that the Tsrielson correlation suffices to show a gap between local and quantum sum-rates over a wide range of channel parameters $p$ and $q$, i.e., the superiority of a quantum sum-rate over a local one. That is, the Tsirelson correlation is sufficient to show that a quantum capacity is higher than its local counterpart. In addition, other quantum correlations are also considered with various points in the quantum set, which gives the same conclusion.

\subsection{ (b).~~Upper bound  \label{apdxVb}}

For deriving an upper bound, we construct a polytope $\mathcal{P}_{out}$ containing the quantum set, i.e., such that $  \mathcal{Q} \subsetneq \mathcal{P}_{out}\subsetneq \mathcal{NS}$. This polytope is constructed as follows, first we find that among all the $204160$ vertices of the $\mathcal{NS}$ polytope there are only eight vertices, all belonging to the set $V_{\frac{1}{2}}$, which gives the maximum possible sum-rate value $2$ for the channel $\N$ with $p=1$ and $q=0$. Let us denote these eight vertices by $\{v_k: k\in \{1,2,3,4,5,6,7,8\}\}$, and the notation $v_k = \{ P(0,0| 0,0), P(0,1| 0,0), \cdots, P(3,3|1,1) \}$ for a vector in the probability space. Then the eight vertices that we find are as follows:
\begin{widetext}
	\begin{eqnarray}
	v_1=\{0, \frac{1}{2}, 0, 0,  \frac{1}{2}, 0, 0, 0, 0, 0, 0, 0, 0, 0, 0, 0,0, 0, 0,  \frac{1}{2}, 0, 0,  \frac{1}{2}, 0, 0, 0, 0, 0, 0, 0, 0, 0, \nonumber \\
	0, 0, 0, 0, 0, 0, 0, 0, 0,  \frac{1}{2}, 0, 0,  \frac{1}{2}, 0, 0, 0,
	0, 0, 0, 0, 0, 0, 0, 0, 0, 0,  \frac{1}{2}, 0, 0, 0, 0,  \frac{1}{2}\}\\
	v_2=\{0, 0, 0, \frac{1}{2}, 0, 0, \frac{1}{2}, 0, 0, 0, 0, 0, 0, 0, 0, 0, 0, \frac{1}{2}, 0, 0, \frac{1}{2}, 0, 0, 0, 0, 0, 0, 0, 0, 0, 0, 0, \nonumber \\
	0, 0, 0, 0, 0, 0, 0, 0, 0, 0, \frac{1}{2}, 0, 0, 0, 0, \frac{1}{2}, 0, 0, 0, 0, 0, 0, 0, 0, 0, \frac{1}{2}, 0, 0, \frac{1}{2}, 0, 0,0\}
	\end{eqnarray}

	\begin{eqnarray}
	v_3=\{0, 0, 0, 0, 0, 0, 0, 0, 0, \frac{1}{2}, 0, 0, \frac{1}{2}, 0, 0, 0, 0, 0, 0, 0, 0, 0, 0, 0, 0, 0, \frac{1}{2}, 0, 0, 0, 0, \frac{1}{2}, \nonumber \\
	0, \frac{1}{2}, 0, 0, \frac{1}{2}, 0, 0, 0, 0, 0, 0, 0, 0, 0, 0, 0, 0, 0, 0, \frac{1}{2}, 0, 0, \frac{1}{2}, 0, 0, 0, 0, 0, 0, 0, 0, 0 \}\\
	v_4=\{0, 0, 0, 0, 0, 0, 0, 0, 0, 0, \frac{1}{2}, 0, 0, 0, 0, \frac{1}{2}, 0, 0, 0, 0, 0, 0, 0, 0, 0, \frac{1}{2}, 0, 0, \frac{1}{2}, 0, 0, 0,  \nonumber \\
	0, 0, 0, \frac{1}{2}, 0, 0,\frac{1}{2}, 0, 0, 0, 0, 0, 0, 0, 0, 0, 0, \frac{1}{2}, 0, 0, \frac{1}{2}, 0, 0, 0, 0, 0, 0, 0, 0, 0, 0, 0 \}
	\end{eqnarray}

	\begin{eqnarray}
	v_5=\{\frac{1}{2}, 0, 0, 0, 0, \frac{1}{2}, 0, 0, 0, 0, 0, 0, 0, 0, 0, 0, 0, 0, \frac{1}{2}, 0, 0, 0, 0, \frac{1}{2}, 0, 0, 0, 0, 0, 0, 0, 0, \nonumber \\
	0, 0, 0, 0, 0, 0, 0, 0, \frac{1}{2}, 0, 0, 0, 0, \frac{1}{2}, 0, 0, 0, 0, 0, 0, 0, 0, 0, 0, 0, 0, 0, \frac{1}{2}, 0, 0, \frac{1}{2}, 0 \}\\
	v_6=\{0, 0, \frac{1}{2}, 0, 0, 0, 0, \frac{1}{2}, 0, 0, 0, 0, 0, 0, 0, 0, \frac{1}{2}, 0, 0, 0, 0,  \frac{1}{2}, 0, 0, 0, 0, 0, 0, 0, 0, 0, 0, \nonumber \\
	0, 0, 0, 0, 0, 0, 0, 0, 0, 0, 0, \frac{1}{2}, 0, 0, \frac{1}{2}, 0, 0, 0, 0, 0, 0, 0, 0, 0, \frac{1}{2}, 0, 0, 0, 0, \frac{1}{2}, 0, 0 \}
	\end{eqnarray}

	\begin{eqnarray}
	v_7=\{0, 0, 0, 0, 0, 0, 0, 0, \frac{1}{2}, 0, 0, 0, 0, \frac{1}{2}, 0, 0, 0, 0, 0, 0, 0, 0, 0, 0, 0, 0, 0, \frac{1}{2}, 0, 0, \frac{1}{2}, 0, \nonumber \\
	\frac{1}{2}, 0, 0, 0, 0, \frac{1}{2}, 0, 0, 0, 0, 0, 0, 0, 0, 0, 0, 0, 0, \frac{1}{2}, 0, 0, 0, 0, \frac{1}{2}, 0, 0, 0, 0, 0, 0, 0, 0 \}\\
	v_8=\{0, 0, 0, 0, 0, 0, 0, 0, 0, 0, 0, \frac{1}{2}, 0, 0, \frac{1}{2}, 0, 0, 0, 0, 0, 0, 0, 0, 0, \frac{1}{2}, 0, 0, 0, 0, \frac{1}{2}, 0, 0, \nonumber \\
	0, 0, \frac{1}{2}, 0, 0, 0, 0, \frac{1}{2}, 0, 0, 0, 0, 0, 0, 0, 0, \frac{1}{2}, 0, 0, 0, 0, \frac{1}{2}, 0, 0, 0, 0, 0, 0, 0, 0, 0, 0 \}.
	\end{eqnarray}
	
\end{widetext}

These eight vertices form four distinct pairs such that any convex combination of two vertices from a pair again gives the maximum possible sum-rate $2$ for the channel with $p=1$ and $q=0$, and the four pairs are $\{v_1,v_2\},\{v_3,v_4\},\{v_5,v_6\},\{v_7,v_8\}$. We also find that on considering a convex combination of any three (or more) out of these eight vertices, except for the case which reduces to convex combination of only two vertices from a pair, the sum rate is strictly less than the maximal achievable value $2$. Since our goal is to construct a polytope $P_{out}$ such that sum-capacity over $\mathcal{P}_{out}$ is less than the sum-capacity over $\mathcal{NS}$ for all possible values of parameters $p$ and $q$, first of all, it is necessary that we remove these eight vertices, as well as, all the other points which are convex combination of two points from any pair, i.e., all points of the $\mathcal{NS}$ polytope which gives the sum rate $2$ for the channel with $p=1$ and $q=0$.  From our construction it will easily follow that $\mathcal{P}_{out}\subsetneq \mathcal{NS}$. Interestingly, moreover we will show that $Q\subsetneq \mathcal{P}_{out}$, and that the constructed polytope $\mathcal{P}_{out}$ will be sufficient for showing a gap between quantum and no-signaling capacities for all the interference channels satisfying $p\neq q$.\\

To construct the polytope $P_{out}$ with the desired property, by looking at the pairs of vertices $\{v_i,v_j\}$ we consider hyperplanes $L_{(i,j)}=c_{ij}$, where $(i,j)\in\{(1,2),(3,4),(5,6), (7,8)\}$, such that these hyperplanes separates quantum set $\mathcal{Q}$ from all points which are convex combination of $\{v_i,v_j\}$. We define these hyperplanes with the help of four linear functional of joint probabilities $P(a, b\vert x, y)$. For this, let us first define a vector of conditional probabilities, of length $64$, as follows 
\begin{equation}
\vec{P}=\{P(a ,b\vert x ,y):  x,y\in\{0,1\},~a,b\in\{0,1,2,3\}\}. \label{probvec}
\end{equation}
Here, in Eq.~(\ref{probvec}), we follow an ordering such that a conditional probability $P(a,b\vert x,y)$ appears at a position computed from expression
\begin{equation}
x~2^5+y~2^4 +a_2~2^3 + a_1~2^2 +b_2~2^1 + b_1 ~2^0 +1, \label{order}
\end{equation}
where $a_1,a_2,b_1,b_2$ are derived from relations $a=a_1a_2$ and $b=b_1b_2$ which are nothing but binary representations of outcomes $a$ and $b$.
We note that ordering of conditional probabilities of any arbitrary point (vector) in the considered CGLMP scenario, we follow the rule for ordering the entries of vectors (or points in the geometric space) as defined by Eqs.~(\ref{probvec}) and (\ref{order}). Then, the four linear functional that we consider can be expressed as follows:
\begin{eqnarray}
L_{(i,j)}&=&2(v_i+v_j)\cdot \vec{P} \label{eq:qbell} \\
(i,j)&\in&\{(1,2),(3,4),(5,6), (7,8)\} \nonumber
\end{eqnarray}

\begin{figure}[t]
	\begin{center}
		\includegraphics[angle=0, width=0.48\textwidth]{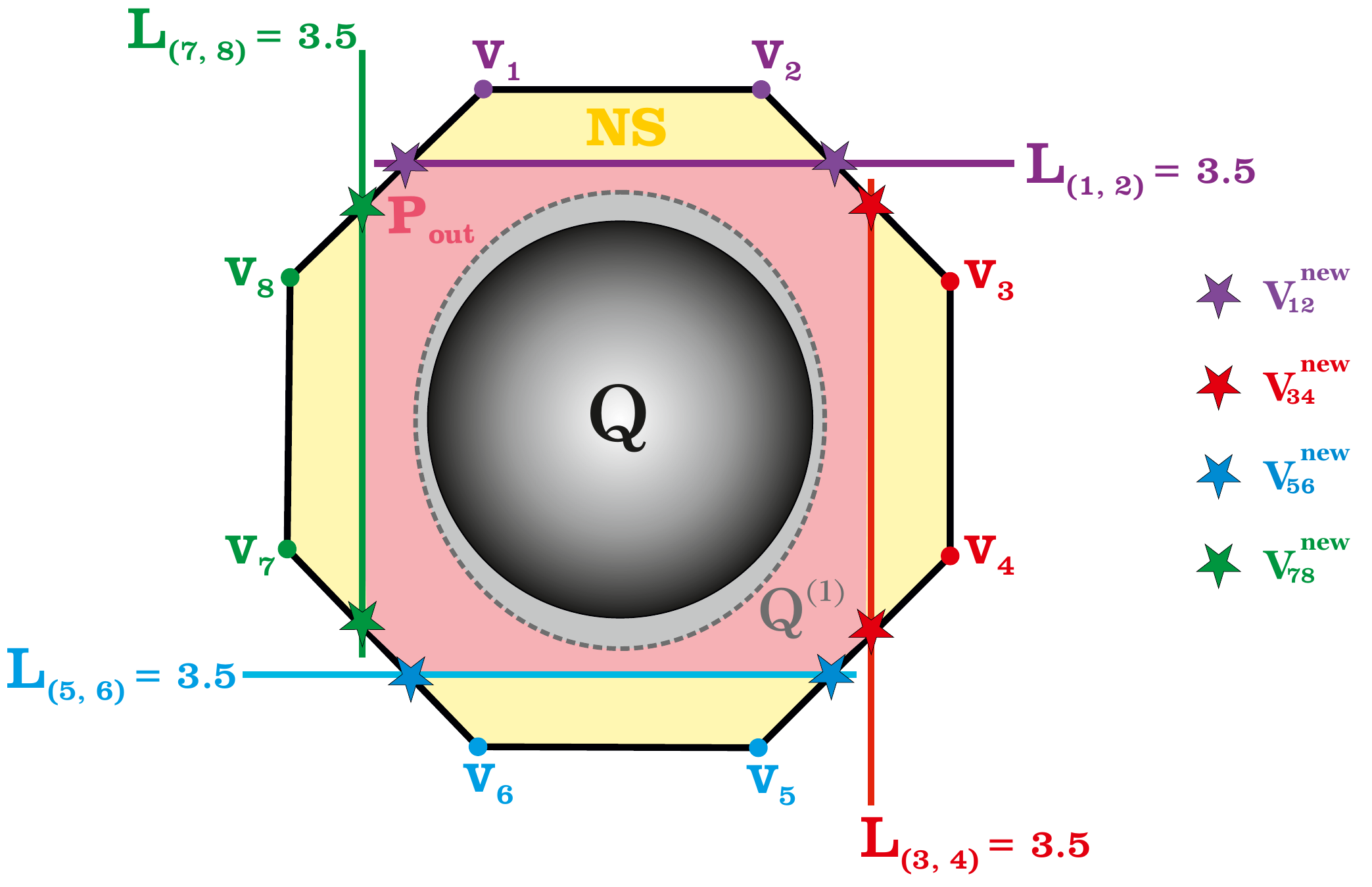}
		\caption{ The set $V_{\frac{1}{2}}$ has $8$ vertices $v_1,\cdots,v_8$ at which the capacity is maximal $ C^{\mathcal{NS}}(\N) =2$. The channel capacity is also maximal on the convex combinations of $(v_1, v_2)$, as well as $(v_3,v_4)$, $(v_5,v_6)$, $(v_7,v_8)$. Since it is clear that these cannot be achieved by quantum resources, they are excluded to find a tighter upper bound to the quantum capacity. \\
			A superset of the set of quantum probabilities is constructed for the purpose, denoted by $P_{out}$, see a polytope in Eq. (\ref{eq:polv}). The quantum set denoted by $Q$ is a subset of $Q^{(1)}$ of the NPA hierarchy, and $P_{out}$ is in fact constructed as a superset of $Q^{(1)}$ as follows. Let $L_{(i,j)}$, see also Eq. (\ref{eq:qbell}), denote a hyperplane that distinguishes a set of non-signaling probabilities that are not in $Q$. A half-space defined by $L_{(i,j)}\leq 3.5$ is satisfied by quantum probabilities. This is a quantum analogy to a Bell inequality: a half-space is satisfied by quantum probabilities but some non-local probabilities. Since it is hard to characterize $Q$, we numerically compute the maximum value of $L_{(i,j)}$ over the probabilities satisfying $Q^{(1)}$, see Eq. (\ref{eq:qbell1}), and obtain about $3.414$. We put $L_{(i,j)} =3.5$ so that the half-space with $L_{(i,j)} \leq 3.5$ includes $Q^{(1)}$. Let $V_{ij}^{new}$ denote the new vertices generated by excluding those probabilities that do not satisfy $L_{(i,j)} \leq 3.5$ in the non-signaling polytope. In this way, vertices violating $L_{(i,j)}\leq 3.5$ for $(i,j) \in \{(1,2), (3,4), (5,6), (7,8) \}$ are excluded in the non-signaling polytope. The set of remaining vertices in the probability space constructs a polytope $P_{out}$ a superset of the quantum set. The vertices are characterized in Eq. (\ref{eq:pol})}  \label{newfigure}
	\end{center}	
\end{figure}

Next for bounding the quantum set with suitable hyperplanes constructed from linear functionals $L_{(i,j)}$, we need to find maximum and minimum value of these functionals over the quantum set $\mathcal{Q}$, which is a convex optimization problem. However, since we do not know the exact quantum set we use a series of outer approximations developed by Navascues-Pironio-Acin (NPA) \cite{NPA1,NPA2} which converges to the quantum set $\mathcal{Q}$. So we choose our domain as different levels of the NPA-hierarchy, denoted here by $Q^{(k)}$ where $k\in\{0,1,\mbox{1+ab}, 2, 3,...\}$. All these different levels are convex, and they form a sequence of outer approximations of the set of quantum correlation $Q$, i.e., $Q^{(0)}\supseteq Q^{(1)}\supseteq ... Q^{(k)}...\supseteq Q$. Thus we solve the convex optimization problem which can be written as a semidefinite program as follows	
\begin{eqnarray}
&~&\mbox{Max}~(\mbox{Min})~\left[ L_{(i,j)}\right] \label{eq:qbell1}\\
&~&\mbox{Subject~to}\nonumber \\
&~&\vec{P}(a,b|x,y) \in Q^{(k)}. \nonumber
\end{eqnarray}
By implementing the function NPAHierarchy$(\vec{P}, k)$ from QETLAB \cite{qetlab}, at the simplest (nontrivial) level $k=1$, the respective minimum and maximum value that all the four linear functional $L_{(i,j)}$ achieve is $0$ and $3.414$ $(\approx 2+ \sqrt{2})$. Therefore, we can consider the hyperplanes $L_{(i,j)}=0$ and $L_{(i,j)}=3.5$ for cropping the no-signaling polytope $\mathcal{NS}$, which will give a polytope $\mathcal{P}_{out}$ containing the quantum set $\mathcal{Q}$. Since the hyperplanes  $L_{(i,j)}=0$ are basically faces of the $\mathcal{NS}$ polytope, thus for obtaining the polytope $\mathcal{P}_{out}$ it is sufficient to consider intersection of half-spaces  $L_{(i,j)}\leq3.5$ with the $\mathcal{NS}$. This in turn will remove some vertices of the $\mathcal{NS}$ polytope and add some new ones. First we note that for  our constructed hyperplanes, by running a simple code on MATHEMATICA, we checked and find that all the points in the set $V_1,V_{\frac{1}{4}},V_{\frac{1}{3}}$, and $\{V_{\frac{1}{2}}-\{v_k: k\in\{1,...,8\}\}$ satisfy $L_{(i,j)}\leq 3.5~~\forall (i,j)\in \{(1,2),(3,4),(5,6),(7,8)\}$, whereas all the eight vertices $\{v_k: k\in\{1,...,8\}\}$ give the value $L_{(i,j)}= 4$, for some $(i,j)\in \{(1,2),(3,4),(5,6),(7,8)\}$, and thus all these eight vertices are outside the quantum set $Q$.  These features, as described above, lead us to determine the set of vertices of the polytope $P_{out}$ as follow 

	\begin{eqnarray}
	V_{P_{out}}&=& [V_{\frac{1}{2}}-\{v_k: k\in\{1,...,8\}]\cup V_{12}^{new}\cup V_{34}^{new}\nonumber \\&\cup&V_{56}^{new} 
	 \cup V_{78}^{new}\cup V_{\frac{1}{3}}\cup V_{\frac{1}{4}}\cup V_{1}, \label{eq:pol}
	\end{eqnarray}

here $V_{ij}^{new}$ represents the new vertices generated by cutting the no-signaling polytope $\mathcal{NS}$ with the hyperplane $L_{(i,j)}=3.5$. The constructed polytope is then given by
\begin{equation}
P_{out}= \mbox{ConvexHull}~\{V_{P_{out}}\}\label{eq:polv}
\end{equation}

In order to compute all the new vertices, we first note that set of new vertices $V_{12}^{new}$ are vertices of another (smaller) polytope defined by all normalization conditions, no-signaling conditions, and the hyperplane $L_{(i,j)}=3.5$. Set of all these linear equations is thus one representation of this polytope, what we need here is to find the vertex representation of this polytope. This is an instance of a standard problem in polyhdral geometry which can be solved exactly by using the software Polymake~\cite{Polymake}, we thus obtained all the vertices $V_{ij}^{new}$, and for each of the four possible $(i,j)$ pairs the number of new vertices are $3070$. 

Now in the end, what remains to check formally is that the polytope $P_{out}$ indeed has the expected properties, for this we state and prove the following proposition: \\

{\bf Proposition.} $\mathcal{Q}\subsetneq P_{out} \subsetneq \mathcal{NS}$.\\

\emph{Proof:} It is easy to see that by construction  $P_{out} \subsetneq \mathcal{NS}$. Next, we show that $\mathcal{Q}\subsetneq P_{out}$. Suppose it is not true, then there exist a point $x$ such that $x\in Q$ and $x \notin P_{out}$. Since $x\in Q$, it satisfies all the linear constraints defining the no-signaling polytope $\mathcal{NS}$. Thus, the only way to satisfy  $x \notin P_{out}$ is to violate $L_{(i,j)}\leq 3.5$ for some $(i,j)\in\{(1,2),(3,4),(5,6), (7,8)\}$ which contradicts our assumption $x\in Q$ due to the fact that $\max_{\mathcal{Q}}L_{(i,j)}\approxeq 3.414$ for all $(i,j)$. Hence $x\in \mathcal{Q}\Rightarrow x\in P_{out}$, and therefore $\mathcal{Q}\subsetneq P_{out}$. \\

Finally for obtaining an upper bound on quantum capacity, similar to the method described for computing the no-signaling capacity, we maximize the sum-rate over all points in $V_{P_{out}}$, i.e., over all the vertices of the polytope $\mathcal{P}_{out}$. First, with the help of MATHEMATICA, we compute the sum-rate $C_{E}^{(Q_{UB})}$ over the set of all vertices $V_{P_{out}}$ of the polytope $P_{out}$, these are basically different functions of $p$ and $q$. Then by noticing that many of the vertices give same sum-rate functions, we delete all the duplicates to reduce to very few number of distinct functions. Since for finding an expression for capacity we need to take maximum over all these functions, we filtered and eliminated all the functions which are always less than or equal to a smaller subset of the set of distinct functions. We find that such smallest subset contains only four different functions of $p$ and $q$. The resulting expression for the obtained upper bound is as follows,
\begin{eqnarray}
&& C^{ (\Q_{UB})} (\N)=\max~\{ g  (p,q), g (q,p), f_{UB}(p,q) , f_{UB}(q,p) \},\nonumber \\
&&\mathrm{where} ~g (p,q) =1+h ( \frac{2+p}{5} )-\frac{1}{2} h (\frac{1+3p}{5} )-\frac{3}{2} h(p),\nonumber \\ 
&& \mathrm{ and}~  f_{UB} (p,q) =2 h ( \frac{5+p-q}{10} ) - h ( \frac{4p+q}{5} ) - h( p ). ~~~~\label{eq:qub}
\end{eqnarray}

\section{VI.~~Three vertices for comparisons  \label{apdxVI}}

To study the quantitative relation between the non-locality and the channel capacity, three vertices of the no-signaling polytope are considered: 
\begin{eqnarray}
v_{\frac{1}{2}}=\{\frac{1}{2}, 0, 0, 0, 0, \frac{1}{2}, 0, 0, 0, 0, 0, 0, 0, 0, 0, 0,  \nonumber \\
0, 0, \frac{1}{2}, 0, 0, 0, 0, \frac{1}{2}, 0, 0, 0, 0, 0, 0, 0, 0, \nonumber \\
0, 0, 0, 0, 0, 0, 0, 0, \frac{1}{2}, 0, 0, 0, 0, \frac{1}{2}, 0, 0, \nonumber \\
0, 0, 0, 0, 0, 0, 0, 0, 0, 0, 0, \frac{1}{2}, 0, 0, \frac{1}{2}, 0 \},
\end{eqnarray}

\begin{eqnarray}             
v_{\frac{1}{3}}=\{\frac{1}{3}, 0, 0, 0, 0, \frac{1}{3}, 0, 0, 0, 0, 0, 0, 0, 0, 0, \frac{1}{3},  \nonumber \\
0, 0, \frac{1}{3}, 0, 0, 0, 0, \frac{1}{3}, 0, 0, 0, 0, \frac{1}{3}, 0, 0, 0, \nonumber \\
0, 0, 0, \frac{1}{3}, 0, 0, 0, 0, \frac{1}{3}, 0, 0, 0, 0, \frac{1}{3}, 0, 0, \nonumber \\
0, 0, \frac{1}{3}, 0, 0, 0, 0, 0, 0, 0, 0, \frac{1}{3}, \frac{1}{3}, 0, 0, 0 \},              
\end{eqnarray}

\begin{eqnarray}                  
v_{\frac{1}{4}}=\{\frac{1}{4}, 0, 0, 0, 0, \frac{1}{4}, 0, 0, 0, 0, \frac{1}{4}, 0, 0, 0, 0, \frac{1}{4},  \nonumber \\
0, 0, \frac{1}{4}, 0, 0, 0, 0, \frac{1}{4}, 0, \frac{1}{4}, 0, 0, \frac{1}{4}, 0, 0, 0, \nonumber \\
0, 0, 0, \frac{1}{4}, 0, 0, \frac{1}{4}, 0, \frac{1}{4}, 0, 0, 0, 0, \frac{1}{4}, 0, 0, \nonumber \\
0, \frac{1}{4}, 0, 0, 0, 0,\frac{1}{4}, 0, 0, 0, 0, \frac{1}{4}, \frac{1}{4}, 0, 0, 0 \}.
\end{eqnarray}
Note that all these violate the $2$-input and $4$-outcome CGLMP inequality. In the main text, it is shown that Bell violations are ordered as
\bea
\widetilde{B}_4 ( v_{\frac{1}{4}}) >   \widetilde{B}_4 ( v_{\frac{1}{3}}) > \widetilde{B}_4 ( v_{\frac{1}{2}}) \nonumber
\eea
whereas the sum rates are structured as
\bea
C_{v_{\frac{1}{2}}}^{(\NL)} (\N) >  C_{v_{\frac{1}{3}}}^{(\NL)} (\N) > C_{v_{\frac{1}{4}}}^{(\NL)} (\N). \nonumber
\eea
It is shown that more non-locality does not necessarily imply a higher channel capacity.

\end{document}